\documentclass[12pt]{iopart}
\usepackage{setstack}
\usepackage{graphicx}
\usepackage{cite}

\begin{document}

\title{Interferometric description of optical metamaterials}

\author{P Grahn, A Shevchenko and M Kaivola}

\address{Department of Applied Physics, Aalto University, P.O. Box 13500, FI-00076 Aalto, Finland}

\ead{patrick.grahn@aalto.fi}

\begin{abstract}
We introduce a simple theoretical model that describes the interaction of light with optical metamaterials in terms of interfering optical plane waves. In this model, a metamaterial is considered to consist of planar arrays of densely packed nanoparticles. In the analysis, each such array reduces to an infinitely thin homogeneous sheet. The transmission and reflection coefficients of this sheet are found to be equal to those of an isolated nanoparticle array and, therefore, they are easy to evaluate numerically for arbitrary shapes and arrangements of the particles. The presented theory enables fast calculation of electromagnetic fields interacting with a metamaterial slab of an arbitrary size, which, for example, can be used to retrieve the effective refractive index and wave impedance in the material. It is also shown to accurately describe optically anisotropic metamaterials that in addition exhibit strong spatial dispersion, such as bifacial metamaterials.
\end{abstract}

\maketitle

Optical metamaterials are man-made materials composed of densely packed subwavelength-size nanoparticles appearing like artificial atoms to light. While the optical response of each individual nanoparticle can be revealed using, e.g., the electromagnetic multipole expansion \cite{Multipole}, the description of the macroscopic optical response of a real three-dimensional metamaterial still remains a challenge. This description is complicated by non-trivial interactions between the nanoparticles, including evanescent-wave coupling between them. In this work, we however propose a way to calculate the transmission and reflection characteristics of metamaterial slabs, including anisotropic and spatially dispersive ones, without resorting to evanescent waves.

So far, both numerical \cite{Obrien02,Chiral10,CoreShell11,Alaee13} and experimental \cite{Pshenay-Severin10,Choi11,Gompf2012,Chen12} techniques have been used to obtain the transmission and reflection coefficients for metamaterial slabs with a rather limited number of nanostructured layers. The coefficients obtained for such slabs do not necessarily describe the properties of a bulk metamaterial. One approach to characterize the material would be to successively increase the number of layers and see if the optical characteristics converge \cite{Chiral10,CoreShell11,Chen12}. However, the understanding of the real physics that determines the final transmission properties is lost when using this procedure. The important question is then whether a certain metamaterial slab can be treated as a slab of homogeneous material. To answer this question, one can for example calculate the Bloch eigenmodes in an infinitely extended metamaterial and then, in an additional calculation, check which modes are involved when light is reflected by a semi-infinite metamaterial \cite{Bloch1,Bloch2}. If the calculations show that all but the fundamental Bloch mode are negligible, one can introduce wave parameters, such as the refractive index and wave impedance, for this mode and treat the material as homogeneous. Here, we propose a more straightforward approach, where the properties of a homogenizable metamaterial slab of any thickness are directly linked to the properties of a single layer of the material.

Recently, several retrieval procedures have been introduced to obtain effective wave parameters, such as refractive index and wave impedance, from the reflection and transmission coefficients of a metamaterial slab \cite{Bloch1,Smith02,Chen04,Smith05,Menzel08,Kwon08,ChiralMeta}. These retrieval procedures rely upon the Fresnel coefficients which are derived for dipolar media. However, for the class of bifacial metamaterials in which the electric quadrupole excitations are present \cite{Metadimer}, the classical electromagnetic boundary conditions do not hold \cite{Graham2000,Graham2001}. Consequently, neither the standard Fresnel coefficients nor the retrieval procedures based on these coefficients can be applied to these metamaterials. The development of an adequate theory for the description of highly spatially dispersive metamaterials, such as bifacial metamaterials, would be of great practical importance, e.g., for solar cell applications.

In this work, we re-examine the propagation of light through a slab of spatially dispersive optical metamaterial. We find that both the transmission and reflection by the slab can be surprisingly accurately described in terms of propagating optical plane waves only, which dramatically simplifies the description. We introduce simple analytical expressions that enable one to evaluate the optical characteristics of a thick metamaterial slab by using only the transmission and reflection coefficients of a single layer of the nanoparticles. The single-layer coefficients are evaluated numerically. These coefficients enable one to calculate the electromagnetic fields inside the material and thereby evaluate the wave parameters, such as the refractive index and wave impedance. Compared to previously reported theoretical approaches to the problem, our approach is easy to use in practice independently of the shapes and material compositions of the nanoparticles and of the propagation direction and polarization of the optical waves as long as the material is homogenizable.

\begin{figure}
\centering
\includegraphics[scale=0.75]{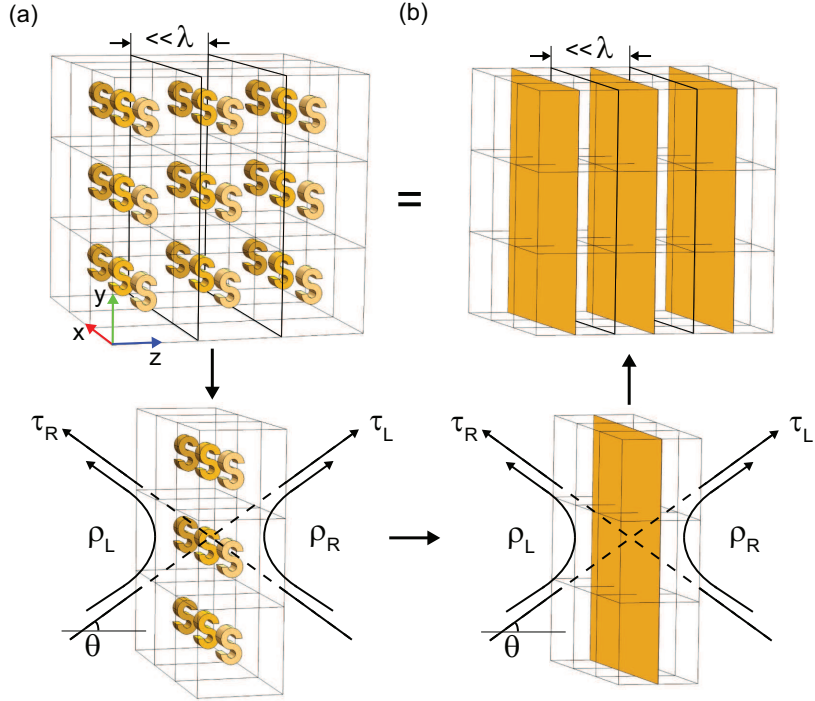}
\caption{A metamaterial slab (a) that is composed by stacking layers of arbitrary scatterers, S, is described as (b) an array of infinitesimally thin sheets. The transmission and reflection coefficients ($\tau_{\rm{L}}$, $\tau_{\rm{R}}$, $\rho_{\rm{L}}$ and $\rho_{\rm{R}}$) for each of these sheets are taken to be equal to those of a single isolated layer of the metamaterial.\label{fig1}}
\end{figure}

Consider a metamaterial slab that is created by stacking two-dimensional arrays of nanoscatterers in a transparent dielectric medium (see figure~\ref{fig1}a). When illuminated by an optical plane wave, the first layer of nanoscatterers transmits a certain portion of the incident field. A part of this transmitted field is then reflected back by the second layer and the rest is partially absorbed and partially transmitted further to the third layer. Provided that the periodicity within each layer is sufficiently smaller than the illumination wavelength, the diffracted waves will be evanescent and the propagating field between the neighboring layers will consist of two optical plane waves. Furthermore, since the particles in each layer are packed very densely, their \emph{collective} evanescent field can have a very short decay length in the direction perpendicular to the layer; note that the \emph{individual} evanescent fields of the particles are still of a long range compared to the particle separation. This effect makes it possible to neglect the near-fields of the layers and describe each layer simply as an infinitesimally thin sheet surrounded by the host dielectric (see figure~\ref{fig1}b). The total transmission and reflection of the slab is then described in terms of the transmission and reflection coefficients of the individual sheets in a way resembling the description of Fabry-Perot interferometers. Similar division of metamaterial slabs into homogenized layers has been applied in the context of effective electromagnetic parameter retrieval \cite{Cabuz07,Bloch12}.

Our approach is as follows. For a single isolated layer of a metamaterial, we first numerically calculate the transmission and reflection coefficients and assign them to an equivalent infinitesimally thin sheet in the middle of the unit cells. These coefficients depend on both the angle of incidence $\theta$ and the polarization of the incident field. In addition, for bifacial nanoscatterer arrays, the reflection coefficient changes if the illumination direction is reversed \cite{Metadimer,Metadimer2}. Therefore, for the wave propagating to the right (left) within the metamaterial, we use the reflection coefficient $\rho_{\rm{L}}$ ($\rho_{\rm{R}}$) to describe the reflection from the left (right) side of each layer. Likewise, the transmission coefficient $\tau_{\rm{L}}$ ($\tau_{\rm{R}}$) describes the transmission of a wave incident from the left (right) side of each layer. Optical reciprocity ensures that the transmission coefficients must be the same if normal incidence illumination is considered. When calculated in this way, the parameters $\rho_{\rm{L}}$, $\rho_{\rm{R}}$, $\tau_{\rm{L}}$ and $\tau_{\rm{R}}$ automatically include the near- and far-field coupling between the scatterers within the layer in question. As has been already mentioned, this coupling makes the extent of the evanescent wave along the layer's normal shorter for denser packing of the scatterers.

\begin{figure}
\includegraphics[scale=1.3]{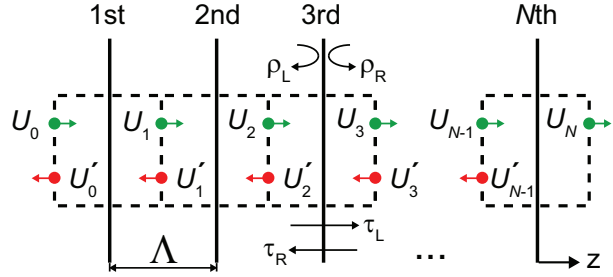}
\centering
\caption{Light propagation through a metamaterial described by an array of infinitesimally thin sheets. Between each pair of such sheets there are two counter-propagating plane waves with transverse field components $U_j$ and $U^{'}_j$. The unit cells of the metamaterial are shown with dashed lines.\label{fig2}}
\end{figure}

Let a plane wave with a transverse field amplitude $U_0$ and a wave vector $\bi{k}_{\rm{in}}=\hat{\bi{x}}k_x+\hat{\bi{z}}k_z$ be incident on a metamaterial slab that consists of $N$ layers of thickness $\Lambda$ and has its surface normal along the $z$ axis (see figure~\ref{fig2}). The wave is assumed to be either TE- or TM-polarized. Treating each nanoscatterer layer as an infinitesimally thin sheet, we consider the counter-propagating waves between the sheets to have wave vectors $\bi{k}_{\pm}=\hat{\bi{x}}k_x\pm\hat{\bi{z}}k_z$, because the material between the sheets is considered to be the same as outside the slab. Assuming that the polarization state is conserved, the transverse fields $U_j$ and $U_j^{'}$ after each sheet $j$ in figure~\ref{fig2} must satisfy the following equations
\begin{equation}\label{rel1}
U_{j} = f_{\rm{L}} U_{j-1}+g_{\rm{R}} U_{j}^{'},
\end{equation}
\begin{equation}\label{rel2}
U_j^{'} = g_{\rm{L}} U_j+f_{\rm{R}} U_{j+1}^{'},
\end{equation}
where $f_{\rm{L}}=\tau_{\rm{L}} \exp(\rmi k_z\Lambda)$, $f_{\rm{R}}=\tau_{\rm{R}} \exp(\rmi k_z\Lambda)$, $g_{\rm{L}}=\rho_{\rm{L}} \exp(\rmi k_z\Lambda)$ and $g_{\rm{R}}=\rho_{\rm{R}} \exp(\rmi k_z\Lambda)$. For chiral metamaterials, both polarization states must be considered simultaneously. In this case, $U_{j}$ and $U_j^{'}$ would be two-element vectors, while $f_{\rm{L}}$, $f_{\rm{R}}$, $g_{\rm{L}}$ and $g_{\rm{R}}$ would be $2\times2$ matrices. Using (\ref{rel1}) and (\ref{rel2}), we derive separate relations for the forward and backward propagating fields
\begin{eqnarray}\label{rec1}
\beta U_{j+1}+U_{j-1}-\alpha U_j = 0,\\\label{rec2}
\beta U_{j+1}^{'}+U_{j-1}^{'}-\alpha U_j^{'} = 0,
\end{eqnarray}
where $\alpha=f_{\rm{R}}+f_{\rm{L}}^{-1}(1-g_{\rm{L}}g_{\rm{R}})$ and $\beta = f_{\rm{R}}/f_{\rm{L}}$. With the help of (\ref{rel1}) and (\ref{rec1}) and the fact that $U_N^{'} = 0$, we obtain the transmission coefficient of the slab to be
\begin{equation}\label{transmission}
t = \frac{U_N}{U_0} = \frac{f_{\rm{L}}}{G_N-\beta f_{\rm{L}}G_{N-1}}.
\end{equation}
Here we have introduced the $G$-polynomial that is calculated as
\begin{eqnarray}
G_{0} = 0,\\
G_{1} = 1,\\
G_{j} = \alpha G_{j-1}-\beta G_{j-2}.
\end{eqnarray}
Similarly, using (\ref{rel1}), (\ref{rel2}) and (\ref{rec2}) we derive the reflection coefficient
\begin{equation}\label{reflection}
r = \frac{U_0^{'}}{U_0} = g_{\rm{L}}f_{\rm{L}}^{-1} G_{N} t.
\end{equation}
Equations~(\ref{transmission}) and (\ref{reflection}) enable direct calculation of the transmission and reflection coefficients of an arbitrarily thick metamaterial in terms of the transmission and reflection coefficients of an \emph{isolated} monolayer of the metamaterial. For $N=1$, equations~(\ref{transmission}) and (\ref{reflection}) correctly yield $t = \tau_{\rm{L}}\exp(\rmi k_z\Lambda)$ and $r = \rho_{\rm{L}} \exp(\rmi k_z\Lambda)$ and for $N=2$ the well-known results for a Fabry-Perot etalon are obtained.

\begin{figure}
\includegraphics[scale=1.3]{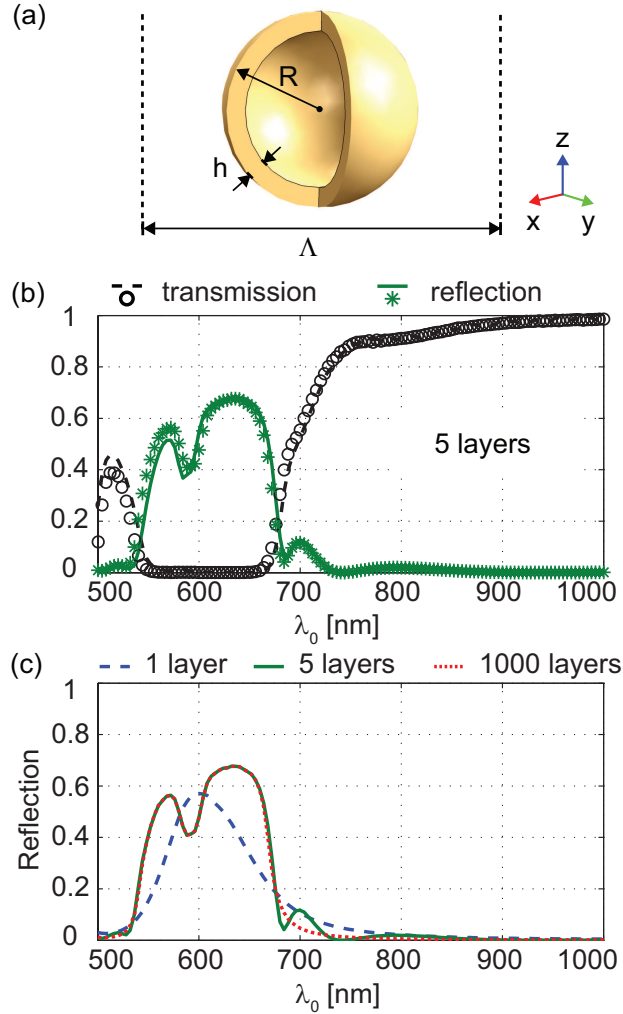}
\centering
\caption{(a) Geometry of the nanoshell [$2R = 70$~nm, $h = 7$~nm, $\Lambda = 130$~nm]. (b) Transmission and reflection spectra of a five-layer thick slab for TM-polarized incident light with the angle of incidence of $45^{\circ}$. The numerically calculated spectra (solid and dashed lines) are shown along with the analytical results (circles and stars) obtained from (\ref{transmission}) and (\ref{reflection}). (c) Reflection spectra for increasing number of layers as obtained from (\ref{transmission}) and (\ref{reflection}).\label{fig3}}
\end{figure}

\begin{figure}
\includegraphics[scale=1.3]{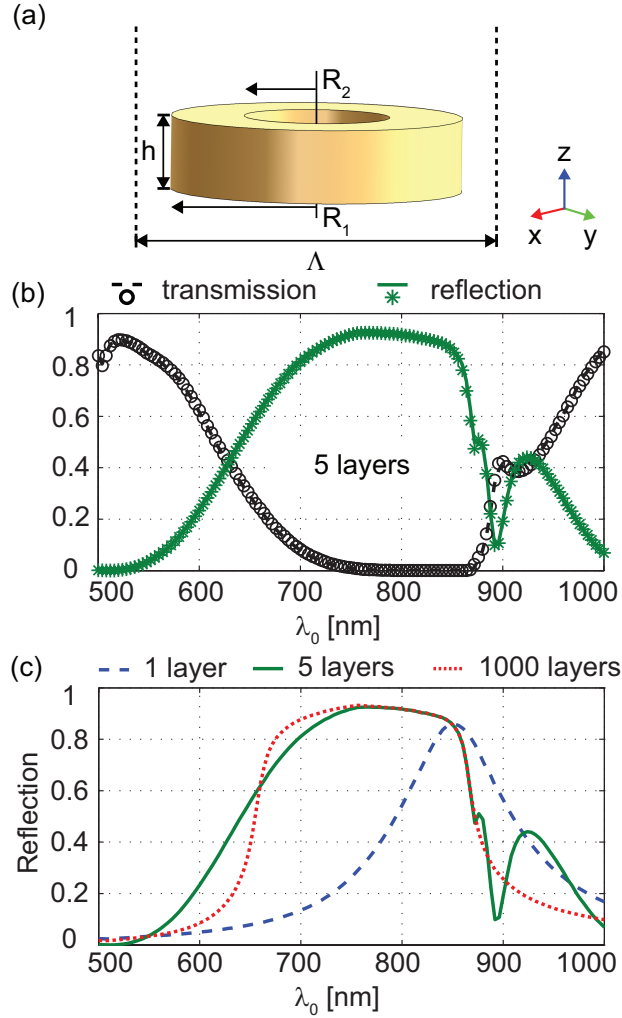}
\centering
\caption{(a) Geometry of the nanoring [$2R_1 = 40$~nm, $2R_2 = 20$~nm, $h = 10$~nm, $\Lambda = 50$~nm]. (b) and (c) are as in figure~\ref{fig3}.\label{fig4}}
\end{figure}

\begin{figure}
\includegraphics[scale=1.3]{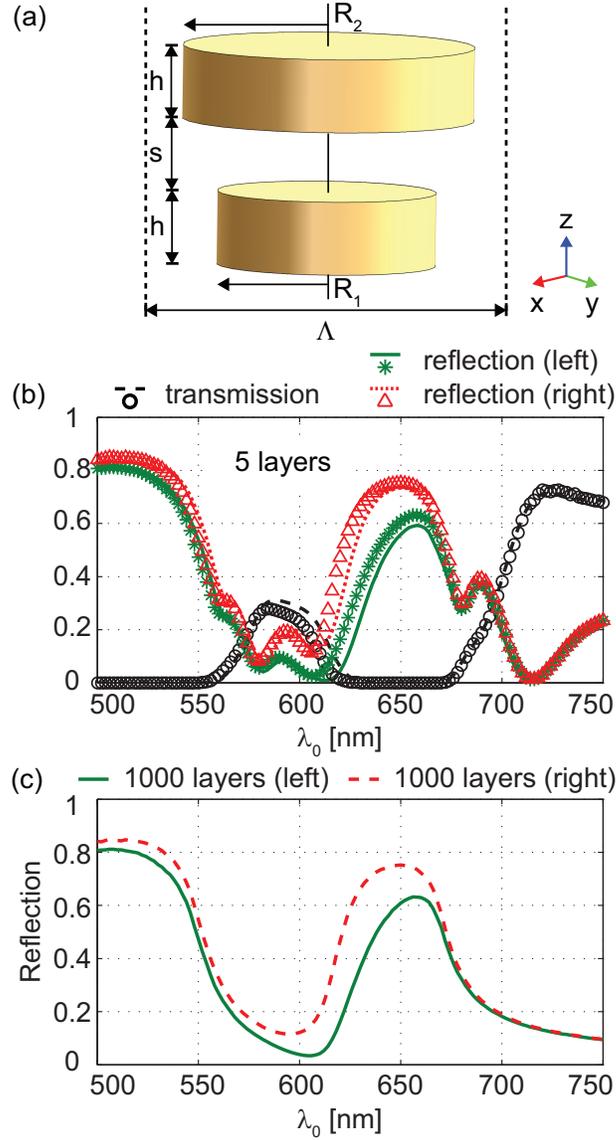}
\centering
\caption{(a) Geometry of the disc nanodimer [$2R_1 = 30$~nm, $2R_2 = 40$~nm, $h = s = 10$~nm, $\Lambda = 50$~nm]. (b) is as in figure~\ref{fig3}b. The numerically and analytically calculated reflections from the side of the larger discs are shown by the additional red dotted line and red triangles, respectively. (c) Reflection spectra for 1000 layers as obtained from (\ref{transmission}) and (\ref{reflection}). \label{fig5}}
\end{figure}

In order to demonstrate the applicability of our theory, we compare it with rigorous numerical calculations. This is done by selecting some non-trivial nanoscatterers, such as nanoshells (figure~\ref{fig3}a), nanorings (figure~\ref{fig4}a) and nanodimers (figure~\ref{fig5}a). These nanoscatterers are considered to compose stacks of two-dimensional periodic arrays that are embedded in a dielectric host medium of refractive index 1.5. The necessary transmission and reflection coefficients for a single array are calculated using the computer software COMSOL Multiphysics. The obtained coefficients are then used in (\ref{transmission}) and (\ref{reflection}) to acquire the transmission and reflection coefficients for several layers. These coefficients are compared with the results of direct numerical calculations of the whole stack with COMSOL. For these calculations we choose TM-polarized illumination with $\theta = 45^{\circ}$ and a slab consisting of $N=5$ layers. This choice is general enough for demonstrating the applicability of the model.

We first consider a metamaterial with an isotropic unit cell containing a silver nanoshell as depicted in figure~\ref{fig3}a. The nanoshells have an outer radius $R = 35$~nm and a thickness $h = 7$~nm. They form a cubic lattice with period $\Lambda = 130$~nm. The optical properties of such nanoshells are well investigated \cite{Shell10,YeWan2011,Vincenti12} and similar structures can be relatively readily fabricated \cite{Yong2006,Prevo2008}. For the calculations, the optical characteristics of silver were taken from \cite{Johnson1972}. The calculated transmission and reflection spectra in the wavelength range from 500 to 1000~nm for the nanoshell slab are shown in figure~\ref{fig3}b. The spectra obtained by using (\ref{transmission}) and (\ref{reflection}) are in a very good agreement with the direct numerical calculations, indicating that the plane-wave description of the light propagation is appropriate and that the evanescent-wave coupling between the layers is indeed weak. A more detailed description of the influence of this coupling on the transmission and reflection spectra is presented later on in the paper.

While direct numerical calculations are always limited by the present computational resources, the introduced theory allows us to calculate the response of an arbitrarily thick slab. In figure~\ref{fig3}c the reflection spectrum for a slab of 1000 nanoshell layers is shown by the red curve. For this particular case of nanoshells, the spectrum of 1000 layers is close to the spectrum of the 5-layer slab and it is already indistinguishable from the spectrum of an infinitely thick metamaterial.

Next, we introduce an anisotropic (uniaxial) unit cell containing a silver nanoring as depicted in figure~\ref{fig4}a. An interesting application of such structures as optical security marks is proposed in \cite{Nanorings}. The ring has an outer radius $R_1 = 20$~nm, inner radius $R_2 = 10$~nm and thickness $h = 10$~nm. The rings form a cubic lattice with period $\Lambda = 50$~nm, such that each layer is aligned with the xy-plane. Note that within each layer the interparticle separation distance is only 10~nm. The calculated transmission and reflection spectra for the nanoring slab are shown in figure~\ref{fig4}b. The theory yields excellent agreement with direct numerical calculations also for these nanoscatterers. The reflection for 1000 layers, depicted by the red curve in figure~\ref{fig4}c, shows that the bulk metamaterial behaves quite differently from a single layer due to the interlayer interaction. One can notice that if the number of layers is large, the metamaterial acts as a spectrally selective broad-band reflector with a nearly flat-top spectral profile.

As a final example, we consider a bifacial metamaterial slab that exhibits strong spatial dispersion. Such metamaterials have not been studied much in terms of their reflection and transmission characteristics. The unit cells of the material contain asymmetric silver nanodimers (see figure~\ref{fig5}a). These nanodimers have been shown to exhibit complete suppression of the electric dipole excitation in a narrow wavelength range when illuminated from the smaller disc side \cite{Metadimer}. However, in a metamaterial, there will be two counter-propagating waves and the electric dipole moment cannot be suppressed for both of them simultaneously \cite{Metadimer2}. The nanodimer geometry is described by the radii $R_1 = 15$~nm and $R_2 = 20$~nm of the discs and dimensions $h = s = 10$~nm defined in figure~\ref{fig5}a. A cubic lattice with period $\Lambda = 50$~nm is now composed of the nanodimers such that the smaller discs are on the left-hand side. As a consequence of the asymmetry of the unit cell, we calculate the single layer response to illumination from both sides in order to obtain the reflection coefficients $\rho_{\rm{L}}$ and $\rho_{\rm{R}}$. The invariance of the nanodimers with respect to rotation around the symmetry axis $z$ ensures that $\tau_{\rm{L}} = \tau_{\rm{R}}$.

Using (\ref{transmission}) and (\ref{reflection}) we calculate the transmission and reflection spectra for a nanodimer slab illuminated from the two sides and compare them with the numerical results. Figure~\ref{fig5}b shows that while the theory very accurately resolves all spectral features, there is a slight deviation of the analytically obtained values from the exact numerical values for the wavelengths around 600~nm. This deviation obviously originates from the evanescent-wave coupling between the adjacent layers. However, considering that the gap size between the discs in the adjacent layers is only 20~nm, the agreement is still remarkably good. We obtained a similar good agreement between our theory and the numerical calculations also for angles of incidence of 0, 30 and 60 degrees as well as for the TE-polarization. The reflection spectra of 1000 layers of nanodimers are depicted in figure~\ref{fig5}c. When illuminated from the small disc side, the reflection coefficient significantly decreases at around the electric dipole suppression wavelength of 618~nm.

Using (\ref{rel1})-(\ref{rec2}), one can retrieve the electric and magnetic fields at any point inside the metamaterial and use them to directly extract the effective wave parameters, such as the refractive index and wave impedance. As an example, consider a non-chiral centrosymmetric material, for which $f_{\rm{L}} = f_{\rm{R}} = f$ and $g_{\rm{L}} = g_{\rm{R}} = g$. Propagation of a plane wave over a single unit cell in the homogenized material must satisfy
\begin{equation}\label{propag}
U_j = U_{j-1}\exp(\rmi \gamma_z\Lambda),
\end{equation}
where $\gamma_z$ is the $z$ component of the effective propagation constant and $\Lambda$ is the unit-cell size in the $z$ direction. Equation (\ref{rec1}) then leads to the following expression
\begin{equation}
\gamma_z\Lambda = \pm\arccos\Big(\frac{1-g^2+f^2}{2f}\Big)+2\pi m,
\end{equation}
where $m$ is an integer. The effective refractive index is related to the wave vector $\bi{k}$ in the host medium through
\begin{equation}
n_{\mathrm{eff}} = \pm \frac{\sqrt{\gamma_z^2+k_x^2}}{k_0},
\end{equation}
which follows from the phase matching condition $k_x = \gamma_x$; $k_0$ is the wave number in vacuum. The effective wave impedance can be obtained by considering the spatially averaged electric and magnetic fields between the sheets introduced in figure~\ref{fig1}. Since in the host medium the fields are right-handed, the total electric field component that is transverse to $z$ is $U_j\exp(\rmi k_z z)+U_j^{'}\exp(-\rmi k_z z)$, whereas the transverse magnetic field component is $[U_j\exp(\rmi k_z z)-U_j^{'}\exp(-\rmi k_z z)][k_z/k]^{p}/Z$, with $p=\pm1$ denoting the TE- and TM-polarizations, respectively. Here $Z$ denotes the wave impedance in the host medium. We can now define
\begin{equation}\label{Zt}
Z_{\perp} = Z\frac{\big<U_j\exp(\rmi k_z z)+U_j^{'}\exp(-\rmi k_z z)\big>}{\big<U_j\exp(\rmi k_z z)-U_j^{'}\exp(-\rmi k_z z)\big>}\Big(\frac{k}{k_z}\Big)^{p}
\end{equation}
that describes the ratio between the transverse components of the averaged electric and magnetic fields; the angle brackets denote averaging over the unit cell. Taking into account the fact that the propagation angle of the effective wave is determined by $\gamma_z/\gamma$, one can find the effective wave impedance $Z_{\rm{eff}} = Z_{\perp}(\gamma_z/\gamma)^p$. Performing the averaging in (\ref{Zt}), we obtain
\begin{equation}
Z_{\rm{eff}} = Z  \frac{g+[1-f\exp(-\rmi \gamma_z\Lambda)]}{g-[1-f\exp(-\rmi \gamma_z\Lambda)]}\Big(\frac{k\gamma_z}{k_z\gamma}\Big)^{p},
\end{equation}
where $U_j^{'}$ was expressed in terms of $U_j$ using (\ref{rel1}) and (\ref{propag}). We recall that $\gamma = n_{\mathrm{eff}}k_0$. One can also obtain the corresponding relative electric permittivity and magnetic permeability as
\begin{equation}
\varepsilon_{\rm{eff}} = \frac{n_{\mathrm{eff}}}{Z_{\rm{eff}}/Z_0},
\end{equation}
\begin{equation}
\mu_{\rm{eff}} = n_{\mathrm{eff}}Z_{\rm{eff}}/Z_0,
\end{equation}
where $Z_0$ is the wave impedance in vacuum. It can be verified that the expressions for $n_{\mathrm{eff}}$, $Z_{\rm{eff}}$, $\varepsilon_{\rm{eff}}$ and $\mu_{\rm{eff}}$ are in full agreement with the commonly used expressions introduced in \cite{Menzel08}. The derivations above can be repeated also for more complex materials, with $f_{\rm{L}} \neq f_{\rm{R}}$ and $g_{\rm{L}} \neq g_{\rm{R}}$.

\begin{figure}
\includegraphics[scale=1.3]{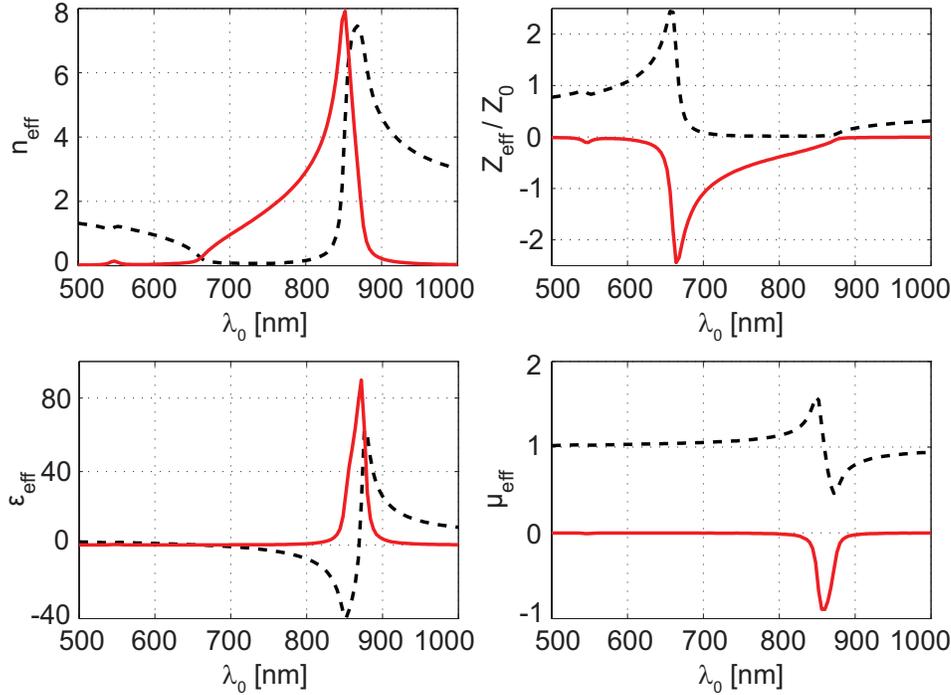}
\centering
\caption{Real (black dashed lines) and imaginary parts (red solid lines) of the effective wave parameters for a material composed of the nanorings shown in figure \ref{fig4}. A TE-polarized wave propagating at $\theta = 45^{\circ}$ in the host medium is considered.\label{fig7}}
\end{figure}

The parameters $n_{\mathrm{eff}}$, $Z_{\rm{eff}}$, $\varepsilon_{\rm{eff}}$ and $\mu_{\rm{eff}}$ calculated for the nanoring material of figure \ref{fig4} are shown in figure~\ref{fig7}, for $\theta = 45^{\circ}$ and TE polarization. This example is of interest in view of the possibility to tune $n_{\mathrm{eff}}$ and $Z_{\rm{eff}}$, because the rings are somewhat similar to traditional split-ring resonators. A strong electric-dipole resonance at around $\lambda_0 = 870$~nm is observed in the spectra of $n_{\mathrm{eff}}$ and $\varepsilon_{\rm{eff}}$. The modification of $\mu_{\rm{eff}}$ in this spectral range is not large and can be interpreted as a result of the finite periodicity $\Lambda$ (see, e.g., \cite{Koschny03}). At wavelengths shorter than $\lambda_0 = 800$~nm, the behavior of $\varepsilon_{\rm{eff}}$ resembles that of a Drude metal. A thick nanoring material could therefore have a high reflectivity in this region, which is supported by the values of the wave impedance. It is interesting, however, that at $\lambda_0 \approx 590$~nm the material is characterized by $n_{\rm{eff}} \approx 1$ and $Z_{\rm{eff}} \approx Z_0$, leading to an efficient suppression of both optical reflection and refraction at an air-metamaterial interface.

The accuracy of the presented theory depends on the extent of the evanescent waves produced by the nanoparticle layers. Qualitatively, for the theory to be exact, the evanescent waves associated with the cut-off diffraction orders must have a decay length that is much shorter than the spacing between the particles in two adjacent layers. For a two-dimensional square array of period $\Lambda_x = \Lambda_y$, the longitudinal wave vector of the first such order $k_{z1} = [k^2-(2\pi/\Lambda_x-k_x)^2]^{1/2}$ is imaginary, with $k$ and $k_x$ being the magnitudes of the total and transverse wave numbers of the incident light. The next order would be $k_{z2} = [k^2-(2\pi/\Lambda_x+k_x)^2]^{1/2}$.  We define the decay length $\delta$ of the evanescent field to be the distance for which the field amplitude has decayed by a factor of $\exp(-1)$. The gap $d$ between the particles in the $z$ direction must then be much larger than
\begin{equation}\label{crit}
\delta = \frac{1}{\mathrm{Im}\{k_{z1}\}} = [(\frac{2\pi}{\Lambda_x}-k_x)^2-k^2]^{-1/2}.
\end{equation}

Considering the nanoshells with $\Lambda_x = 130$~nm, $\lambda_0 = 500$~nm and $k_x = k/\sqrt{2}$, we obtain a decay length $\delta \approx 34$~nm that is smaller than the 60~nm gap between the adjacent shells. For the nanodimers, with $\Lambda_x = 50$~nm, $\lambda_0 = 500$~nm and $k_x = k/\sqrt{2}$, the decay length is $\delta \approx 9$~nm. This value is smaller than the gap $d = 20$~nm between the nanodimers, which supports the success of our analytical calculations. In fact, if for subwavelength-sized unit cells we have $k << 2\pi/\Lambda_x$, (\ref{crit}) yields $\delta \approx \Lambda_x/(2\pi)$. In this case the criterion for neglecting the interlayer evanescent-wave coupling becomes $d >> \Lambda_x/(2\pi)$. Then, as a practical criterion for when our theory can be applied, we require that $d > \Lambda_x/2$.

\begin{figure}
\includegraphics[scale=1.3]{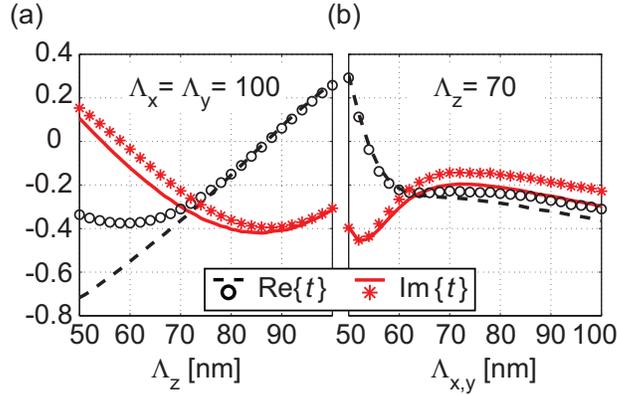}
\centering
\caption{Transmission coefficient $t$ of a five-layer thick nanodimer slab for (a) increasing period $\Lambda_z$ with fixed $\Lambda_x=\Lambda_y=100$~nm and (b) increasing $\Lambda_x=\Lambda_y$ with fixed $\Lambda_z=70$~nm. A TM-polarized illumination with $\lambda_0 = 600$~nm and $\theta = 45^{\circ}$ (from smaller disc side) is considered. The numerically calculated results (solid and dashed lines) are shown along with the analytical results (circles and stars) obtained from (\ref{transmission}) and (\ref{reflection}).\label{fig6}}
\end{figure}

In order to verify the above predictions on the influence of the interlayer evanescent-wave coupling, we numerically calculate the transmission through the nanodimer slab, while varying the transverse and longitudinal periods separately. In figure~\ref{fig6}a the transmission coefficient is plotted for an increasing longitudinal period $\Lambda_z$. The transverse period is fixed to $\Lambda_x = 100$~nm in order to have the evanescent-wave coupling significant enough to cause a discrepancy at small $\Lambda_z$ between the theory and the numerical results. We notice that when $d$ exceeds $\Lambda_x/2$ [$\Lambda_z$ exceeds 80~nm], this discrepancy disappears. The transmission coefficient for an increasing transverse period $\Lambda_x = \Lambda_y$ is shown in figure~\ref{fig6}b. In this case the longitudinal period is fixed to $\Lambda_z = 70$~nm, such that the evanescent-wave coupling is negligible at $\Lambda_x = 50$~nm. As the transverse period is increased, the discrepancy between the theory and the numerical results starts to appear due to an increase in the evanescent-wave coupling between the layers. We notice however that as $\Lambda_x$ is increased, the array also gets sparse, which reduces the influence of the nanoscatterers on the propagating wave. This effect counterbalances the growing decay length of the evanescent waves produced by the nanoparticle layers, such that the discrepancy between the theory and the numerical results remains small.

For a metamaterial slab, in which the evanescent wave coupling is negligible, one can introduce effective material parameters. On the other hand, if the evanescent wave coupling exists, these parameters depend on the slab thickness and are thereby senseless \cite{Simovski}. By using (\ref{transmission}) and (\ref{reflection}) to compare the transmission and reflection coefficients of a single nanoscatterer layer with those of two layers, one can directly assess whether the metamaterial is homogenizable and, consequently, whether the introduction of material parameters is justified.

In summary, we have introduced a simple analytical theory for the description of light interaction with optical metamaterials. Recognizing the subwavelength size of the metamaterial's unit cells, we found that the evanescent-wave coupling between adjacent monolayers of the metamaterial does not significantly influence the light propagation in the material. As a consequence, arbitrarily thick metamaterial slabs can be accurately described in terms of the plane-wave transmission characteristics obtained for a single isolated monolayer, e.g., numerically. Furthermore, we have shown that one can evaluate the fields at any point inside the material and, consequently, obtain the effective wave parameters. The presented examples of rigorous numerical calculations demonstrate the wide applicability of this remarkably simple analytical model.

In contrast to existing theoretical approaches, our one also correctly describes three-dimensional arrays of bifacial nanoscatterers, which is of practical importance for a large variety of metamaterials, such as those with asymmetric unit cells. For homogenizable metamaterial slabs, our method enables rapid one-layer-based extraction of the transmission and reflection coefficients. Furthermore, propagation of an optical beam through such a metamaterial can be described by using the angular spectrum representation with our model applied to each plane-wave component.

The introduced theory is not limited to optical metamaterials, but can also be applied to study wave propagation in other artificial media, such as radio-frequency and terahertz metamaterials, and even phononic metamaterials. We believe that the presented theory has the necessary simplicity and accuracy to accelerate the development of optical metamaterials tailored for real applications.

\ack
This work was funded by the Academy of Finland (Project No. 134029).

\end{document}